\documentclass[3p,times]{elsarticle}
\usepackage{ecrc}

\volume{00}

\firstpage{1}

\journalname{Nuclear Physics A}

\runauth{A. Beraudo}

\jid{nupha}

\usepackage{amssymb}
\usepackage{graphicx}
\usepackage{amsmath}
\newcommand{\beq}{\begin{equation}}
\newcommand{\eeq}{\end{equation}}
\newcommand{\beqa}{\begin{eqnarray}}
\newcommand{\eeqa}{\end{eqnarray}}
\newcommand{\bseq}{\begin{subequations}}
\newcommand{\eseq}{\end{subequations}}

\def\x{{\boldsymbol x}}

\def\0{{\boldsymbol 0}}

\def\simle{\mathrel{\rlap{\raise 0.511ex \hbox{$<$}}{\lower 0.511ex 
\hbox{$\sim$}}}}
\def\simge{\mathrel{ \rlap{\raise 0.511ex 
\hbox{$>$}}{\lower 0.511ex \hbox{$\sim$}}}}

\usepackage[figuresright]{rotating}

\begin{document}

\begin{frontmatter}


\title{Perturbative versus non-perturbative aspects of jet quenching: in-medium breaking of color coherence}

\author{A. Beraudo}

\address{Physics Department, Theory Unit, CERN, CH-1211 Gen\`eve 23, Switzerland}

\begin{abstract}
The quenching of jets (and high-$p_T$ particle spectra) observed in heavy-ion collisions is interpreted as due to the energy lost by hard partons crossing the Quark Gluon Plasma. Here we review recent efforts to include in its modeling important qualitative features of QCD, like the correlations in multiple gluon emissions and the color-flow pattern in parton branchings. In particular, the modification of color connections among the partons of a shower developing in the presence of a medium is a generic occurrence accompanying parton energy-loss. We show how this effect can leave its fingerprints at the hadronization stage, leading by itself to a softening of hadron spectra and to an enhanced production of soft particles in jet-fragmentation. 
\end{abstract}


\end{frontmatter}

\section{Introduction}
The observation of jet-quenching (in a broad sense) represents a strong signature of the formation of an opaque medium in ultra-relativistic nucleus-nucleus collisions, leading to a sizable energy degradation of the high-$p_T$ partons initially produced in hard pQCD events and traversing the fireball. Experimental measurements, initially limited to the suppression of high-$p_T$ hadron production~\cite{PHENIX,STAR,ALICE}, are nowadays available also for the quenching of reconstructed jets~\cite{ATLAS,CMS1,CMS2} in the dense environment produced in heavy-ion collisions.

\begin{figure}[!h]
\begin{center}
\includegraphics[clip,height=3.8cm]{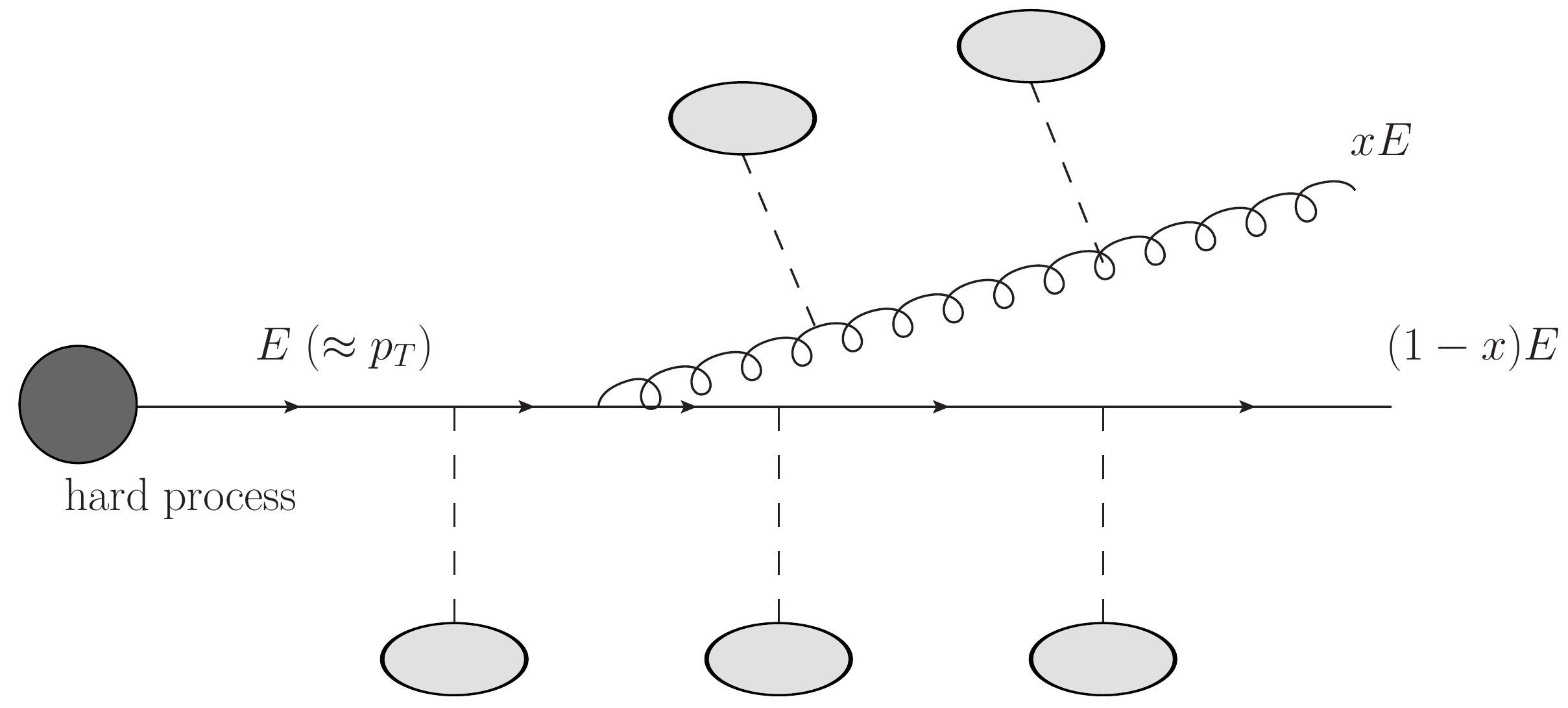}
\caption{Schematic cartoon of parton energy loss due to medium-induced gluon radiation.}
\label{fig:picture}
\end{center}
\end{figure}

The standard theoretical interpretation views the phenomenon as due to parton energy-loss: the interaction of a high energy quark or gluon (produced in a hard pQCD process) with the color field of the medium induces the radiation of (mostly) soft ($\omega\ll E$) and collinear ($k_\perp\ll\omega$) gluons; the emitted gluons, carrying color charge, can further rescatter in the medium, accumulating an additional $q_\perp$ which favors their kinematic decoherence. A schematic picture is given in Fig.~\ref{fig:picture}. The above transverse momentum broadening was shown to be of relevance to explain dijet-imbalance measurements by ATLAS (and CMS) in terms of the out-of cone radiation of many soft partons~\cite{gui}.      

Within the above picture, several models were developed in order to provide a consistent explanation of the experimental data collected at RHIC and LHC. We refer the reader to Ref.~\cite{brick} for a quantitative comparison of the various models available in the literature. Here we prefer to put the emphasis on the conceptual framework they are based on, suggesting how they can be improved in order to account for important qualitative QCD effects which get modified in the presence of a medium. 

If one ignores possible modifications of the PDFs (which will be small for high-$p_T$ observables), the inclusive cross-section for the production of a hadron $X$ in nucleus-nucleus collisions is generally assumed to be described by the following factorized expression
\beqa
d\sigma_{\rm med}^{AA\to h+X}&=&\sum_f d\sigma_{\rm vac}^{AA\to f+X}\otimes{\langle D_{\rm med}^{f\to h}(z,\mu_F^2)\rangle_{AA}}\nonumber\\
{}&=&\sum_f \underbrace{d\sigma_{\rm vac}^{AA\to f+X}}_{pQCD}\otimes\underbrace{{\langle P(\Delta E)\rangle_{AA}}}_{{\rm e.loss\,prob.}} \otimes \underbrace{{D_{\rm vac}^{f\to h}(z,\mu_F^2)}}_{{\rm vacuum\, FF}},\label{eq:fact}
\eeqa
where the medium-modified Fragmentation Function (FF) arises simply from the action of a vacuum FF on a final-state parton having suffered some in-medium energy loss according to the probability density $P(\Delta E)$.

The above factorized expression ignores the role of color correlations, whose effect will be the subject of our presentation.
Firs of all, most calculations of medium-induced gluon radiation focused so far limited on the evaluation of the single inclusive gluon spectrum. Multiple gluon emission was in general described by a simple poissonian distribution
\beq
P(\Delta E)=\sum_{n=0}^\infty\frac{e^{-\langle N_g\rangle}}{n!}\prod_{i=1}^n\left[\int d\omega_i \frac{dN_g}{d\omega_i}\right]\delta\left(\Delta E-\sum_{i=1}^{n}\omega_i\right),
\eeq
treating successive emissions as uncorrelated. 
Secondly, the use of a vacuum FF to describe the transition from partons to hadrons is simply justified through kinematic considerations: the time required for a high-energy parton of virtuality $Q$ to hadronize, of order $Q^{-1}$ in its rest frame, is Lorentz boosted in the lab-frame to ${\Delta t_{\rm lab}^{\rm hadr}}\!\sim\!({E}/{Q})({1}/{Q})$, much larger than the plasma lifetime in the $E\!\to\!\infty$ limit. However the color-exchange interactions of the hard parton with the medium can dramatically change the color connections in the developing shower and this may lead to an effect in the final hadron spectra, independently on when hadronization occurs.
Correlations between successive gluon emissions (leading to the \emph{angular-ordering}~\cite{angord} of vacuum showers) and the color-flow pattern of QCD processes, essential for our understanding of strong interactions in the vacuum, were not accounted for in previous ``jet-quenching'' studies. Here we recall, first, how they provide an essential guidance to interpret the experimental data in elementary collisions. We then discuss their effect in the nucleus-nucleus case.
 
\begin{figure}[!h]
\begin{center}
\includegraphics[clip,height=5cm]{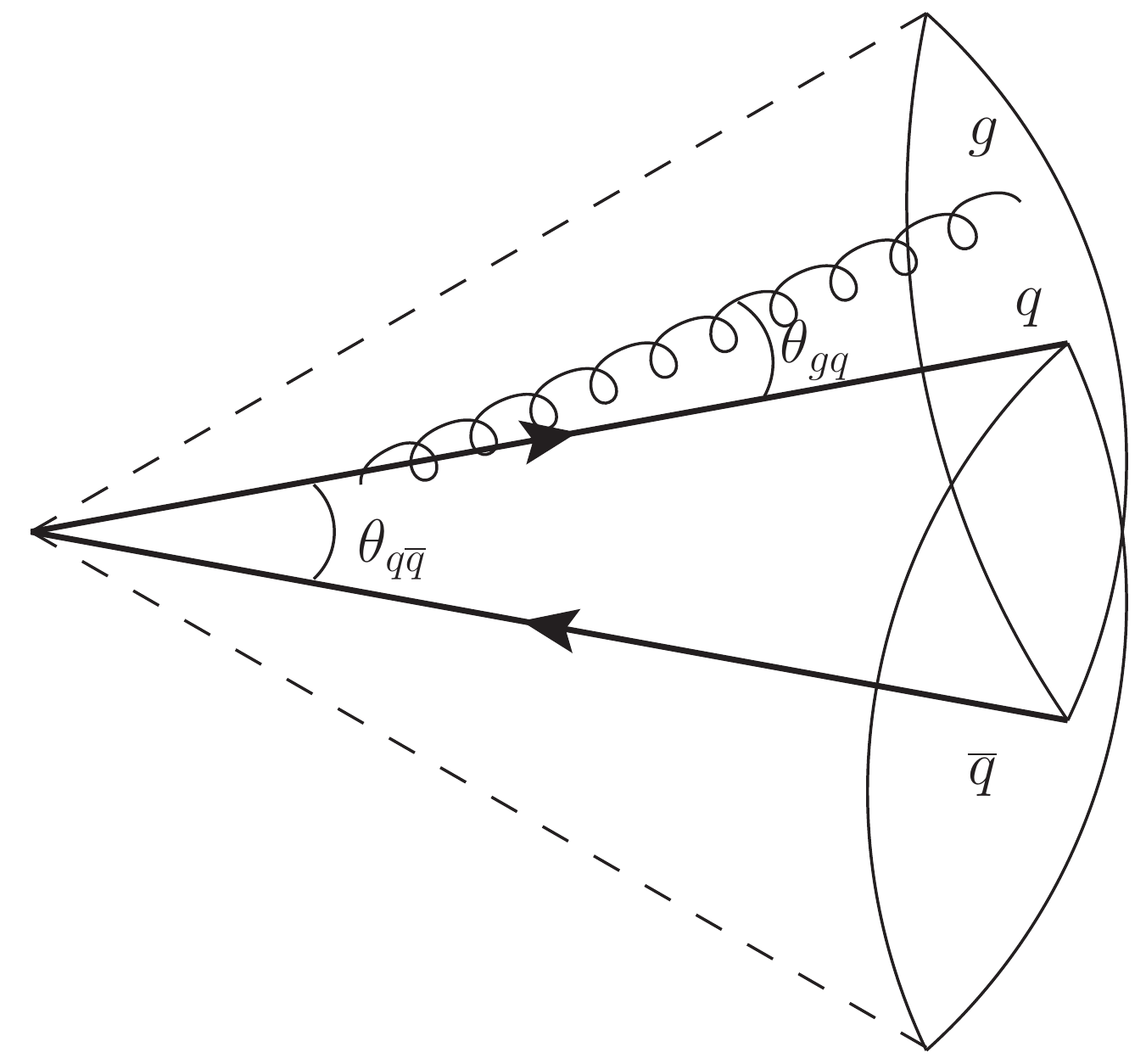}
 \hskip 2cm
\includegraphics[clip,height=5cm]{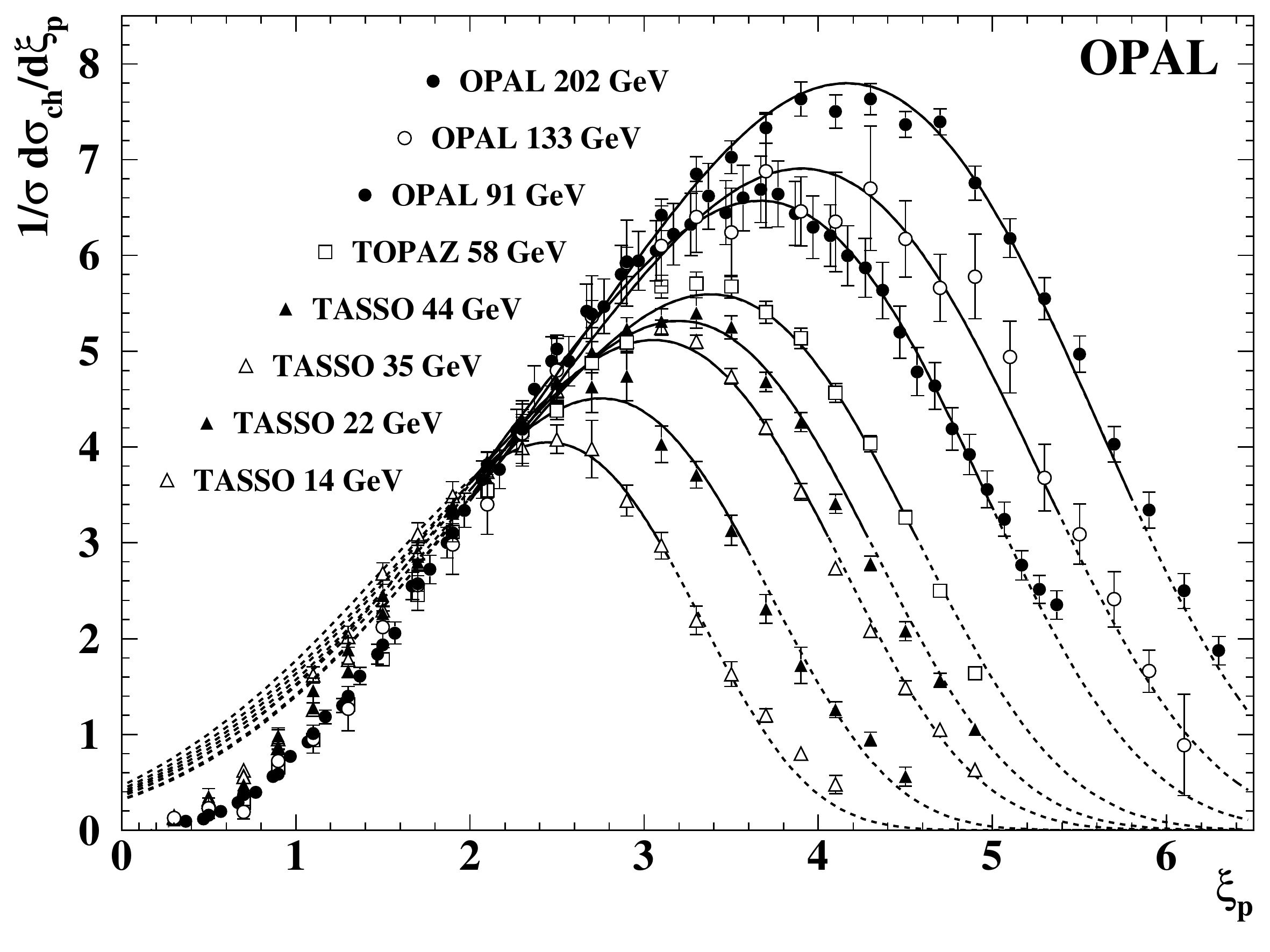}
\caption{Left panel: the angular pattern of the radiation of a color-singlet $q\overline{q}$ antenna. Right panel: the $\xi$ distribution of hadrons from jet-fragmentation, with its characteristic suppression of soft particle production.}
\label{fig:angorder}
\end{center}
\end{figure}
Let us start from the angular distribution of gluon radiation in the vacuum, considering for simplicity the emission from a color-singlet $q\overline{q}$ pair (left panel of Fig.~\ref{fig:angorder}). The $q$ and the $\overline{q}$ form an \emph{antenna} which radiates coherently. During the gluon formation time $t_f\!=\!2\omega/k_\perp^2$ the pair must in fact have reached a separation $d_\perp\!=\!t_f\theta_{q\overline{q}}$ larger than the transverse wave-length $\lambda_\perp\!\sim\!1/k_\perp\!\sim\! 1/\omega\theta_{gq}$ of the gluon, so that the latter can resolve the individual color charges (instead of seeing a neutral object). The request $\lambda_\perp< d_\perp$ forces then the gluon to be radiated within the cone $\theta_{gq}<\theta_{q\overline{q}}$.
Such an \emph{angular ordering} of soft gluon emission in the vacuum leads to important phenomenological consequences. It is at the basis of the development of well collimated jets in hard pQCD events. Furthermore it also explains how hadrons are distributed inside a jet (\emph{intra-jet} coherence). The ordering condition $\lambda_\perp< d_\perp$, entailing $1/k_\perp<(2\omega/k_\perp^2)\,\theta_{q\overline{q}}$, together with the constraint $k_\perp\simge\Lambda_{\rm QCD}$, leads to a lower bound for the energy of the radiated gluon $\omega\simge \Lambda_{\rm QCD}/\theta_{q\overline{q}}$. The angular-ordering condition of QCD radiation was found to nicely accommodate the experimental data -- see the right panel of Fig.~\ref{fig:angorder} displaying the $\xi\!\equiv\!-\ln\left({p^h}/{E^{\rm jet}}\right)$ distribution of hadrons inside a jet~\cite{opal}, with its characteristic \emph{hump-backed plateau}~\cite{hb} -- accounting in particular for the rapid depletion of soft-particle production.   

\begin{figure}[!h]
\begin{center}
\includegraphics[clip,height=4cm]{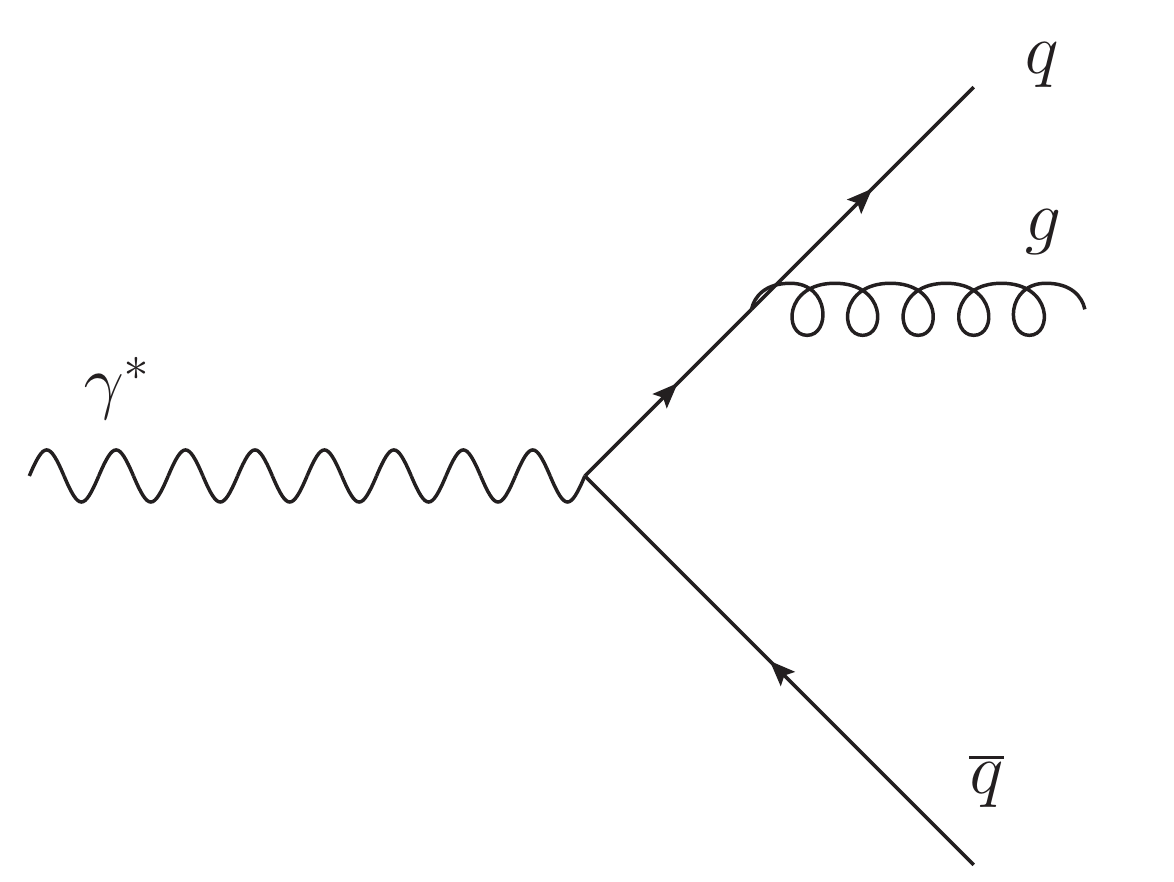}
 \hskip 2cm
\includegraphics[clip,height=4cm]{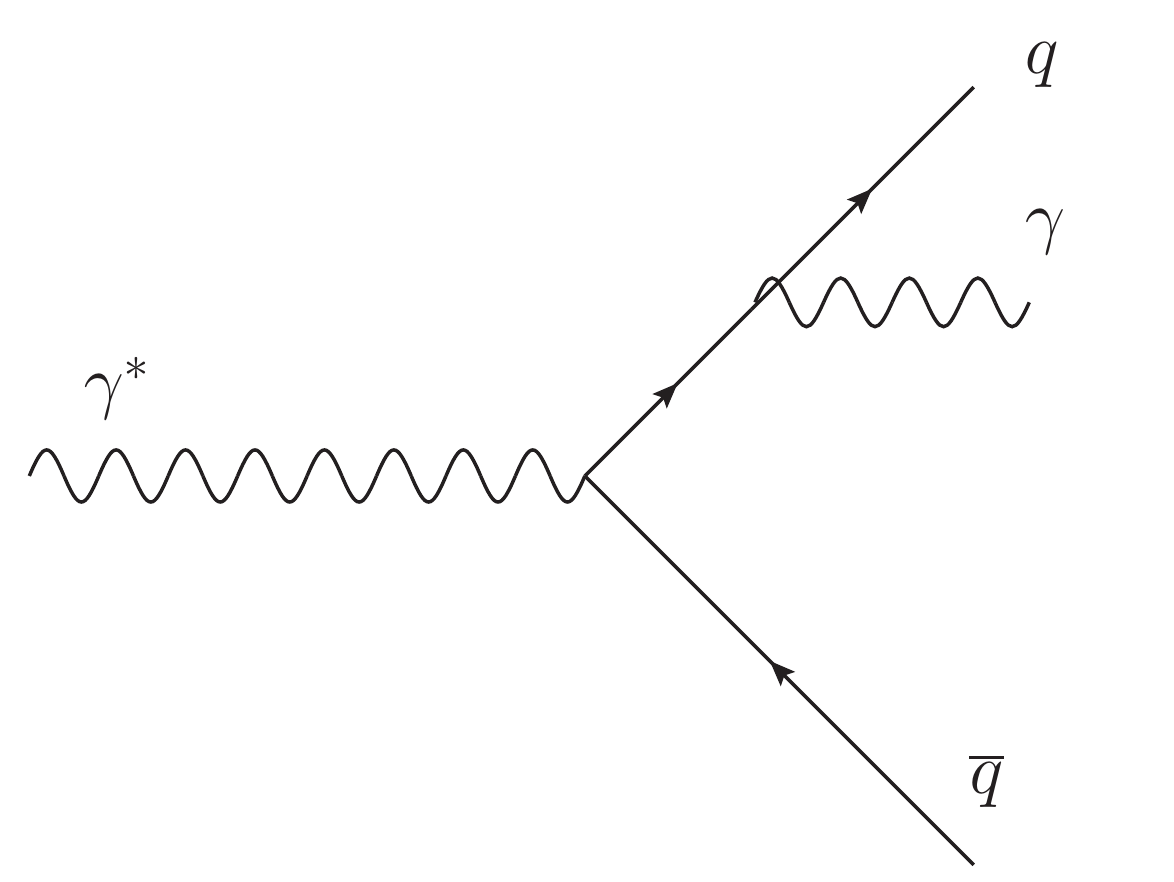}
\includegraphics[clip,height=2.7cm]{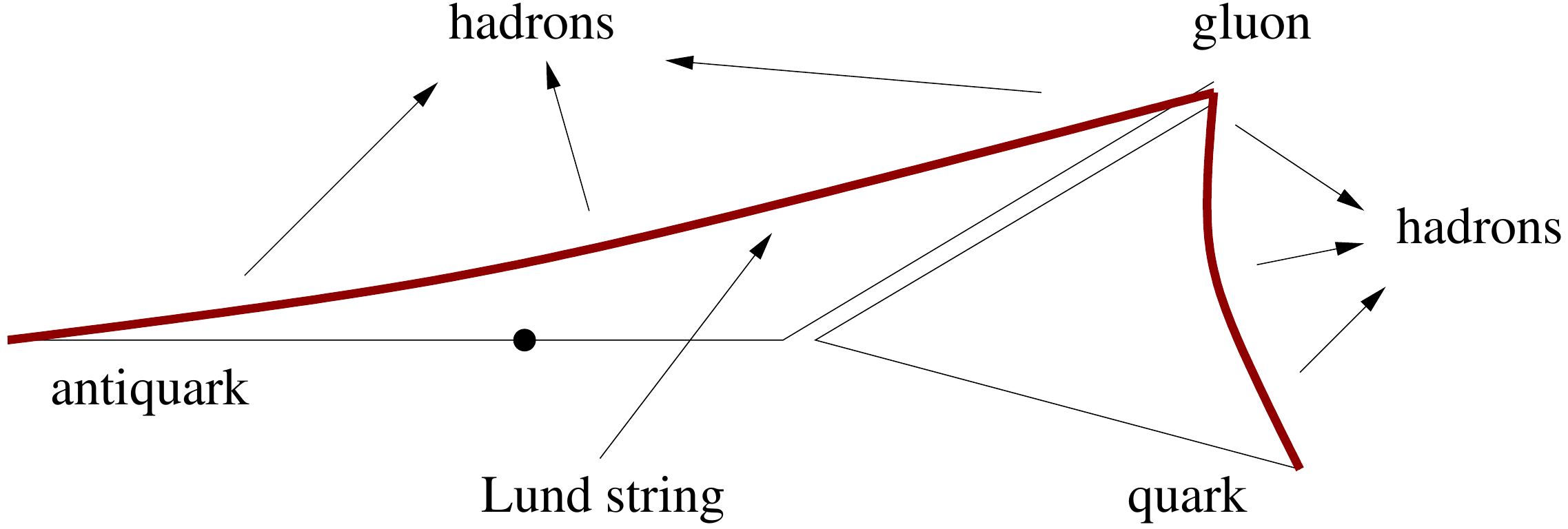}
\hskip 2cm
\includegraphics[clip,height=2.7cm]{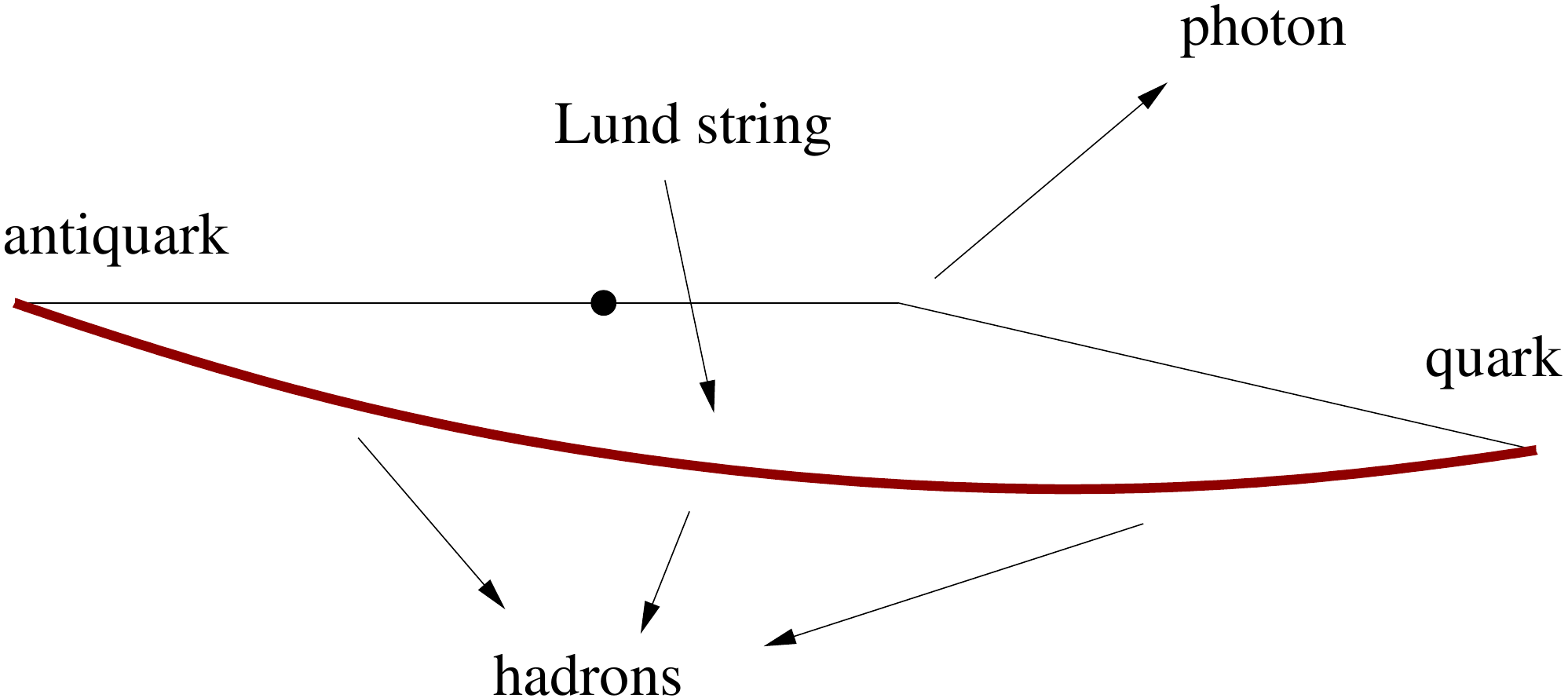}
\caption{Left panel: an $e^+e^-\to q\overline{q}g$ event, with the associated color flow leading to a production of soft particles concentrated in the $q-g$ and $\overline{q}-g$ regions and depleted within the $q-\overline{q}$ angle. Right panel: an $e^+e^-\to q\overline{q}\gamma$ event. With the same kinematics, due to the different color-flow, the angular distribution of soft hadrons is concentrated between the $q$ and the $\overline{q}$ jets.}
\label{fig:string}
\end{center}
\end{figure}
We now consider the role of color-flow in explaining particle production in QCD processes, displaying its relevance in the cleanest possible environment, i.e. $e^+e^-$ collisions. In Fig.~\ref{fig:string} we show two different events -- $e^+e^-\!\to\! q\overline{q}g$ (left panel) and $e^+e^-\!\to\! q\overline{q}\gamma$ (right panel) -- sharing the same kinematics, but having a different color-flow. The final hadrons, here viewed according to the Lund model as coming from the fragmentation of a $q\overline{q}$ string, are then expected to display a different angular distribution in the two cases, with a depletion/enhancement of particle production within the $q\!-\!\overline{q}$ region in the $q\overline{q}g/q\overline{q}\gamma$ cases (\emph{inter-jet} coherence, a.k.a. \emph{string effect}~\cite{string}). Such an effect was actually observed at LEP, see Ref.~\cite{delphi} for more details. Attempts to identify color-coherence effects even in hadronic collisions were also performed~\cite{CDF}, looking for instance at the broader $\eta$-distribution of soft jets due to their color-correlations with initial-state partons moving along the beam axis.

Note that following the color-flow in the large-$N_c$ limit allows to provide also a complementary partonic explanation of inter-jet coherence. For instance, in the $q\overline{q}g$ event in the left panel of Fig.~\ref{fig:string}, further soft gluons would be emitted by the $qg$ and $\overline{q}g$ antennas, radiation within the $q\!-\!\overline{q}$ region being suppressed by a factor $1/N_c^2$. A smooth transition between a perturbative (coherent soft gluon radiation) to a non-perturbative (string fragmentation) description of soft particle production is then possible in the above QCD framework.      

In summary: the angular ordering and color flow of QCD radiation in the vacuum are at the basis of essential phenomena observed in elementary collisions like the development of collimated jets, the hump-backed plateau of the hadron distribution from their fragmentation and the inter-jet coherence of soft particle production. How the above features of QCD radiation are modified by medium effects looks then an interesting issue to address.  

\section{Color decoherence of in-medium gluon radiation}
We now discuss how the antenna radiation and the color flow of QCD processes discussed in the previous section are modified in the presence of a medium. We will simply illustrate the essential qualitative picture, referring the reader to the original literature for all technical details.  
\subsection{Anti-angular ordering}
As reminded in the previous section, in the development of a shower quarks and gluons do not radiate incoherently, but according to antennas formed by pairs of color-connected partons. A major recent achievement has been the evaluation of the radiation of a QCD antenna in the presence of a medium~\cite{yac1,yac2}. In the case of the color-singlet $q\overline{q}$ pair shown in Fig.~\ref{fig:angorder} the radiation spectrum (vacuum + medium-induced contributions) from the quark line (an analogous expression holding for the $\overline{q}$ emission) reads in the soft-limit
\beq
dN_{q,\gamma^*}^{\rm tot}=\frac{\alpha_sC_F}{\pi}{\frac{d\omega}{\omega}}\frac{\sin\theta\,d\theta}{1-\cos\theta}\left[{\theta(\cos\theta-\cos\theta_{q\overline{q}})}+{\Delta_{\rm med}\theta(\cos\theta_{q\overline{q}}-\cos\theta)}\right],
\eeq
where $\Delta_{\rm med}$ is a parameter (to calculate) which, depending on the opacity of the medium, goes from 0 (no medium effect) to 1 (complete decoherence of the $q\overline{q}$ pair). Notice that the medium-induced radiation lies always outside the $q\overline{q}$ cones and that in the limit $\Delta_{\rm med}\!\to\!1$ gluons are emitted without any angular constraints, the $q$ and the $\overline{q}$ radiating as two uncorrelated color charges. Furthermore in such a limit one has also $dN_{\gamma^*}^{\rm tot}|_{\rm opaque}=dN_{g^*}^{\rm tot}|_{\rm opaque}$: the $q\overline{q}$ pair forgets about its initial color charge and the radiation spectrum in the singlet/octet channels looks the same.     

\begin{figure}[!h]
\begin{center}
\includegraphics[clip,height=4.8cm]{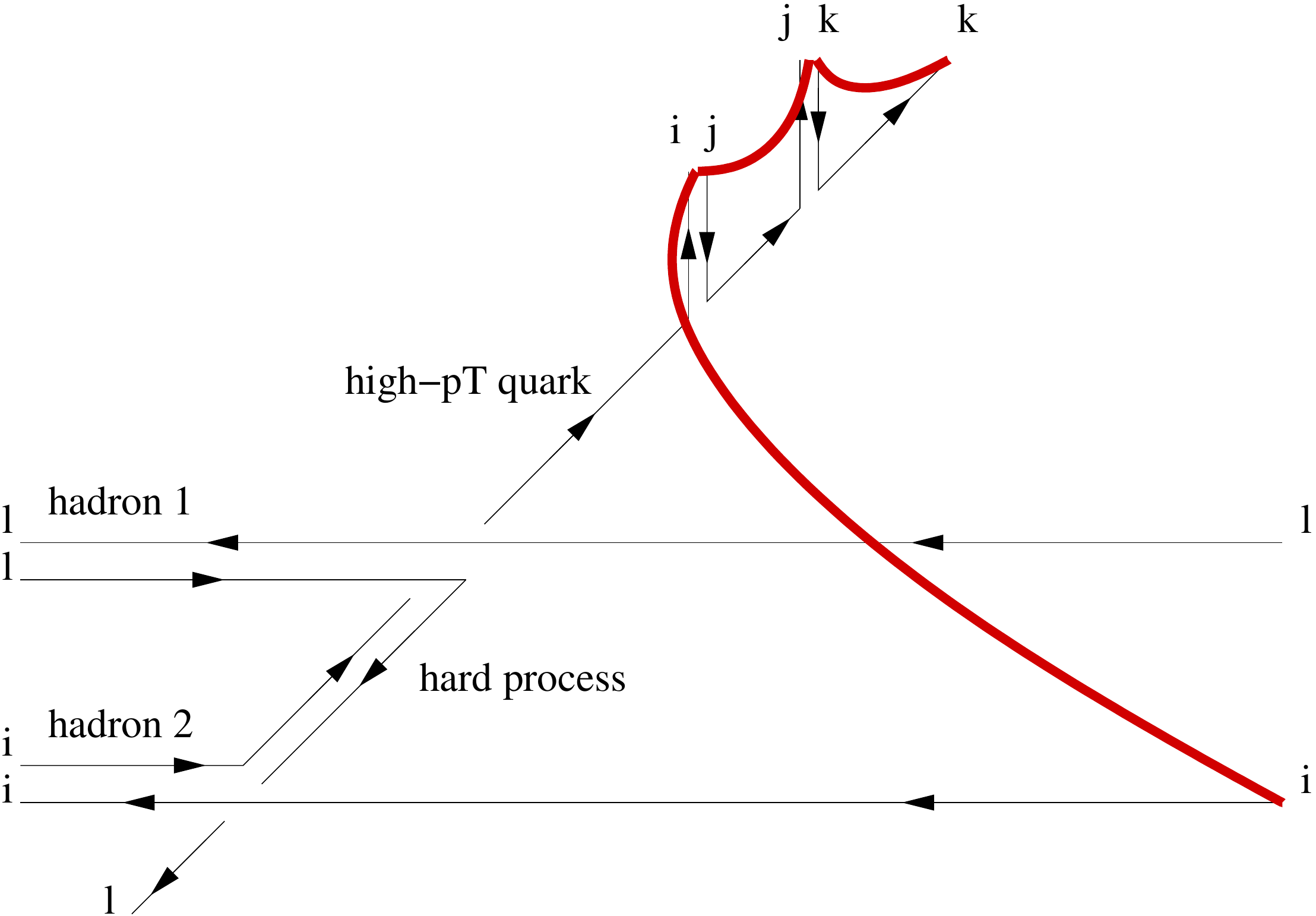}
 \hskip 2cm
\includegraphics[clip,height=4.8cm]{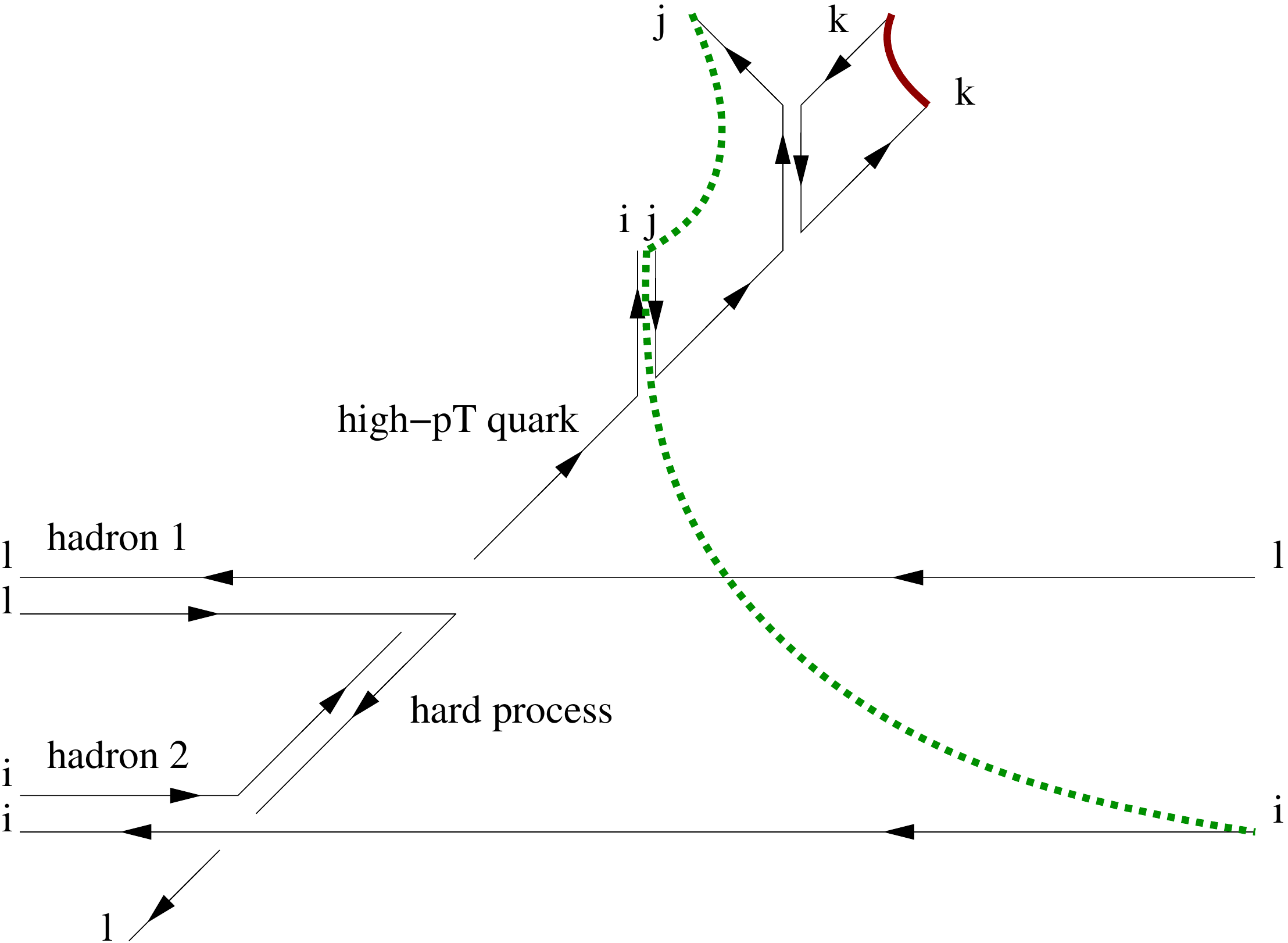}
\caption{Left panel: the typical color flow in a hard event + parton shower in a hadronic collision; the radiated gluons remain color correlated with the leading high-$p_T$ fragment. Right panel: a $g\to q\overline{q}$ splitting may lead to the color decoherence of the shower, but the process -- having no soft-enhancement -- doesn't occur frequently.}
\label{fig:vacshow}
\end{center}
\end{figure}
\subsection{Color flow}
We now focus on the modification of color-flow in parton branchings due to the interaction with a medium and to its possible phenomenological consequences for hadron spectra and jet fragmentation in $AA$ collisions.

The typical color flow of a ``hard event + parton shower'' embedded in an elementary hadronic collision is shown in Fig.~\ref{fig:vacshow}. Colliding hadrons are schematically described as two opposite color charges moving along the beam direction, gluons (according to the large-$N_c$ approximation) are treated as $q\overline{q}$ pairs and final hadrons are supposed to arise from the fragmentation of Lund strings stretching from a quark to an antiquark of the same color. Note that in most of the cases the gluons radiated in the shower are color connected with the leading partonic fragment, being part of the same string (see left panel of Fig.~\ref{fig:vacshow}): if sufficiently collinear they don't entail any increase in the final particle multiplicity and their energy can still be found in the leading hadron (in this sense the vacuum fragmentation pattern can be considered quite hard). The only possibility to break the color connections in the shower is that a gluon undergoes a $g\!\to\!q\overline{q}$ splitting, as shown in the right panel of Fig.~\ref{fig:vacshow}: in this case all the gluons previously radiated are color decohered from the leading fragment and hadronized independently, being part of a separate string. This however is not a process occurring so likely, since the corresponding splitting function $P_{q\leftarrow g}(z)\sim[z^2+(1-z)^2]$ has no soft ($z\!\to\!0$) enhancement as compared for instance with $P_{g\leftarrow q}(z)\sim[1+(1-z)^2]/z$.
    
\begin{figure}[!h]
\begin{center}
\includegraphics[clip,height=5cm]{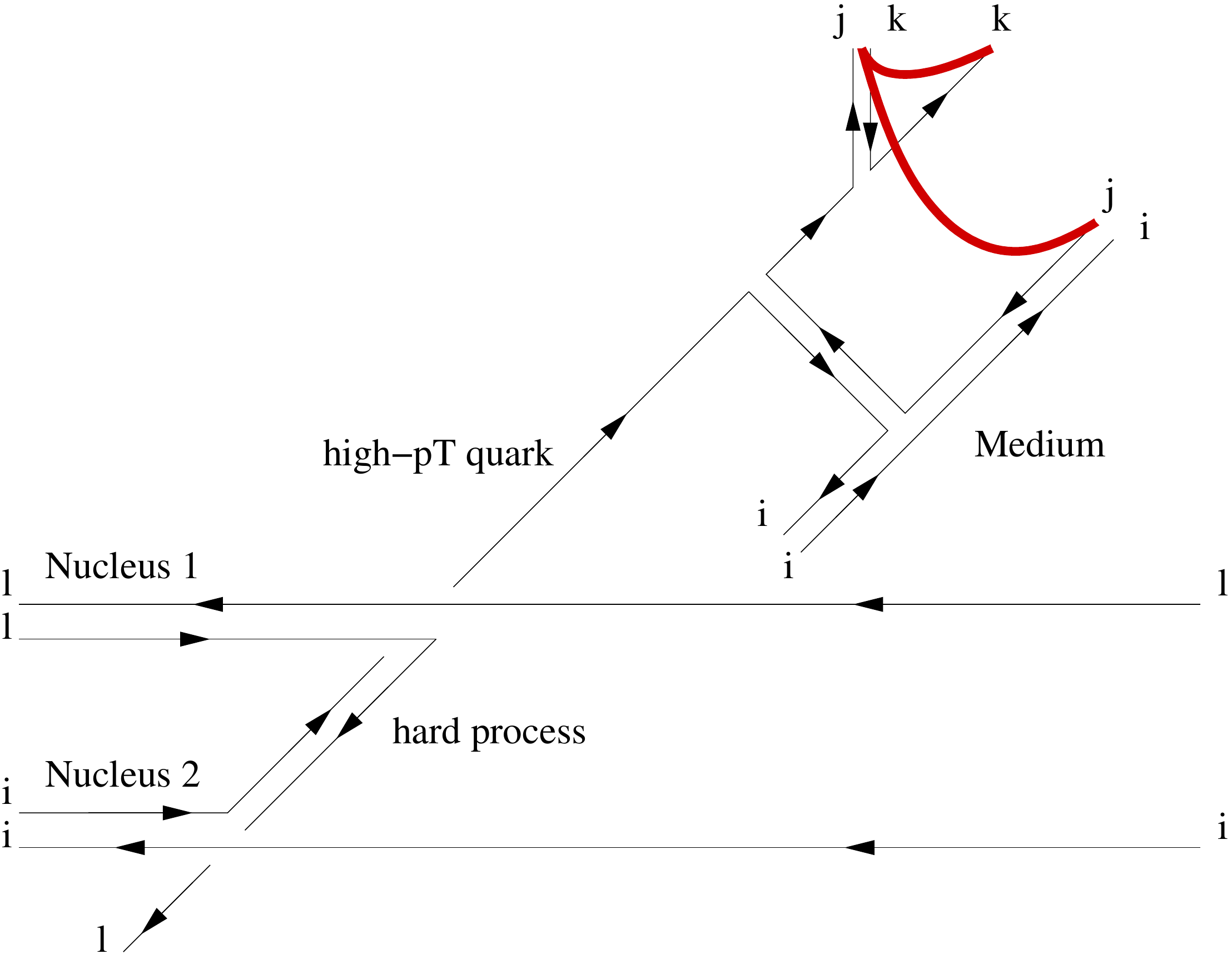}
 \hskip 2cm
\includegraphics[clip,height=5cm]{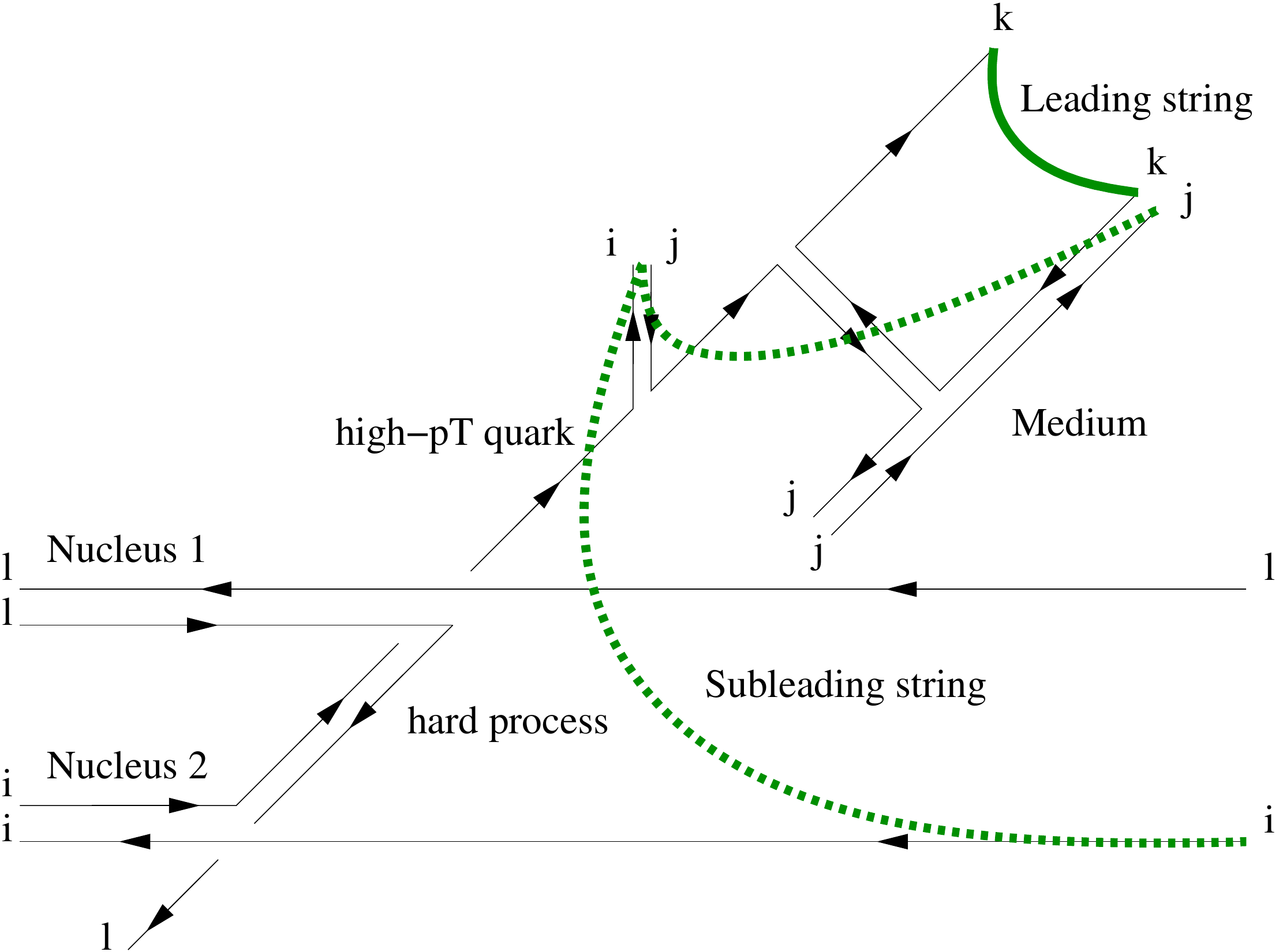}
\caption{The two independent color channels for gluon radiation off a high-$p_T$ parton induced by a single color-exchange with the medium formed in a nucleus-nucleus collision. Left panel: ``Final State Radiation'', with the radiated gluon color connected to the leading fragment. Right panel: ``Initial State Radiation'', with the gluon decohered from the projectile and giving rise to an independent softer string to hadronize.}
\label{fig:medshow}
\end{center}
\end{figure}
Let us now move to nucleus-nucleus collisions. In this case high-$p_T$ partons, produced in hard events, have to cross a dense medium. The rescatterings in its color field induce then the radiation of additional gluons (on top of the usual vacuum emission getting rid of the initial virtuality) leading to an energy-loss of the hard parton. This mechanism alone, however, would require a very dense medium to lead to a sizable suppression of the \emph{hadron} spectra. As we saw discussing vacuum showers, sufficiently collinear gluons just produce small kinks on the string stretched by the high-$p_T$ parton produced in the hard event, without affecting too much the final hadron spectra. In the presence of a medium, from one side, the radiated gluons (carrying color charge) can suffer further rescattering, accumulating additional transverse momentum which favors their \emph{kinematic} decoherence. Here we wish to stress the role of a second source of decoherence, related to the in-medium color rotation of partons (a mechanism stressed also in~\cite{iancu}). As displayed in Fig.~\ref{fig:medshow} (referring for simplicity to the case of a single elastic scattering in the plasma), the interaction of the high-$p_T$ parton with the medium inevitably affects the color connections of the emitted gluons. It is then of interest to setup a color-differential calculation of gluon radiation~\cite{ber1,ber2}, since -- depending on the color flow -- the transition from partons to hadrons can proceed differently. In the left panel of Fig.~\ref{fig:medshow}, labeled as Final State Radiation (FSR) channel, the radiated gluon remains color-connected with the leading partonic fragment and hadronization proceed as in a vacuum shower. On the contrary, in the right panel of Fig.~\ref{fig:medshow}, labeled as Initial State Radiation (ISR) channel, the situation is different. The color connection between the leading fragment and the emitted gluon is broken by the interaction with the medium and the latter is no longer part of the string stretched by the hard parton: its energy is lost and this leads to a softening of the final hadron distribution. Furthermore the decohered gluon is now part of a subleading string, which will be hadronized independently and will contribute to an enhanced multiplicity of soft particles. Note that this second case is actually representative of what is expected to occur more likely in heavy-ion collisions, when multiple elastic collisions in the medium are allowed.   

\begin{figure}[!h]
\begin{center}
\includegraphics[clip,height=5.5cm]{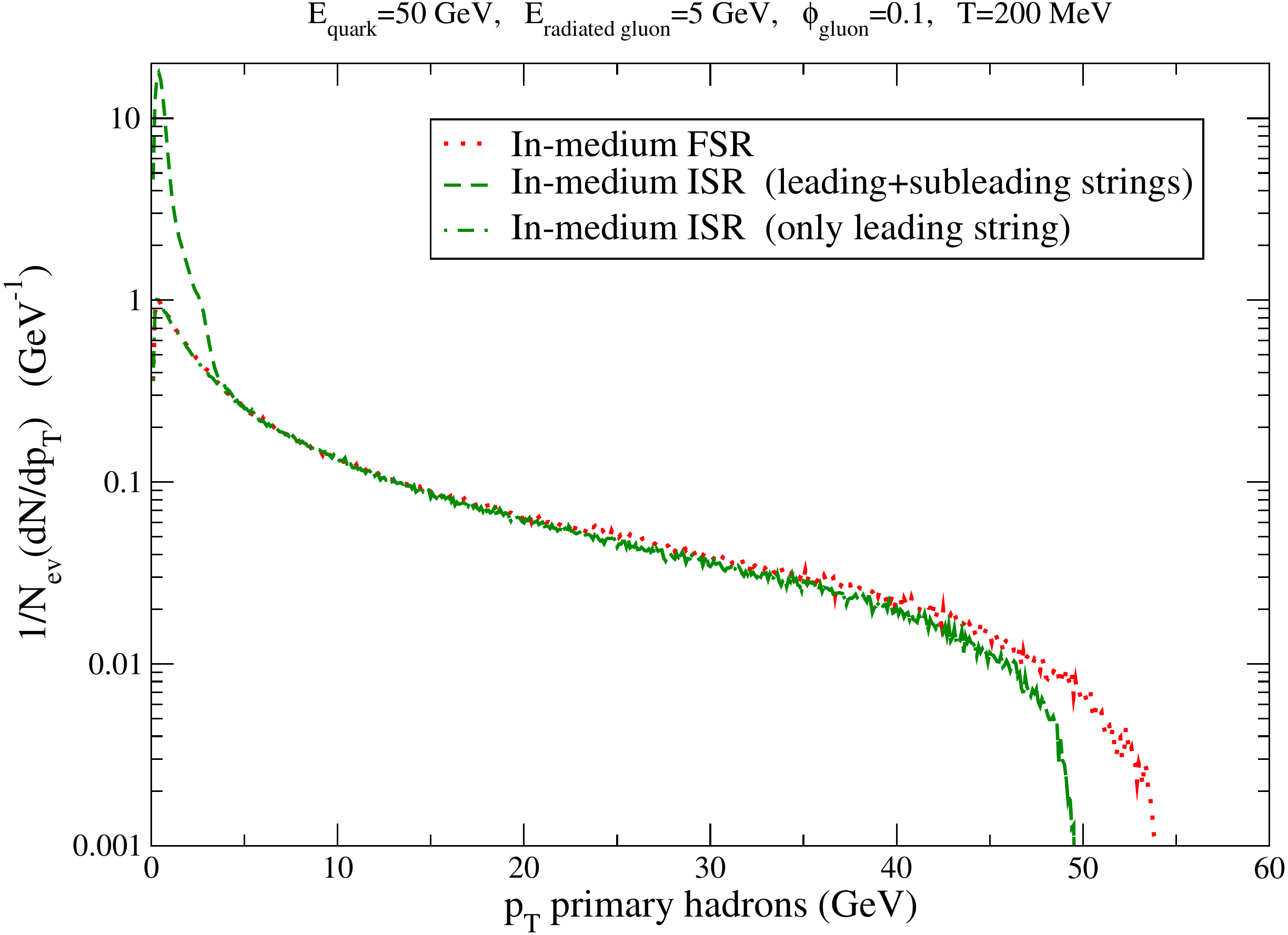}
\hskip 1cm
\includegraphics[clip,height=5.5cm]{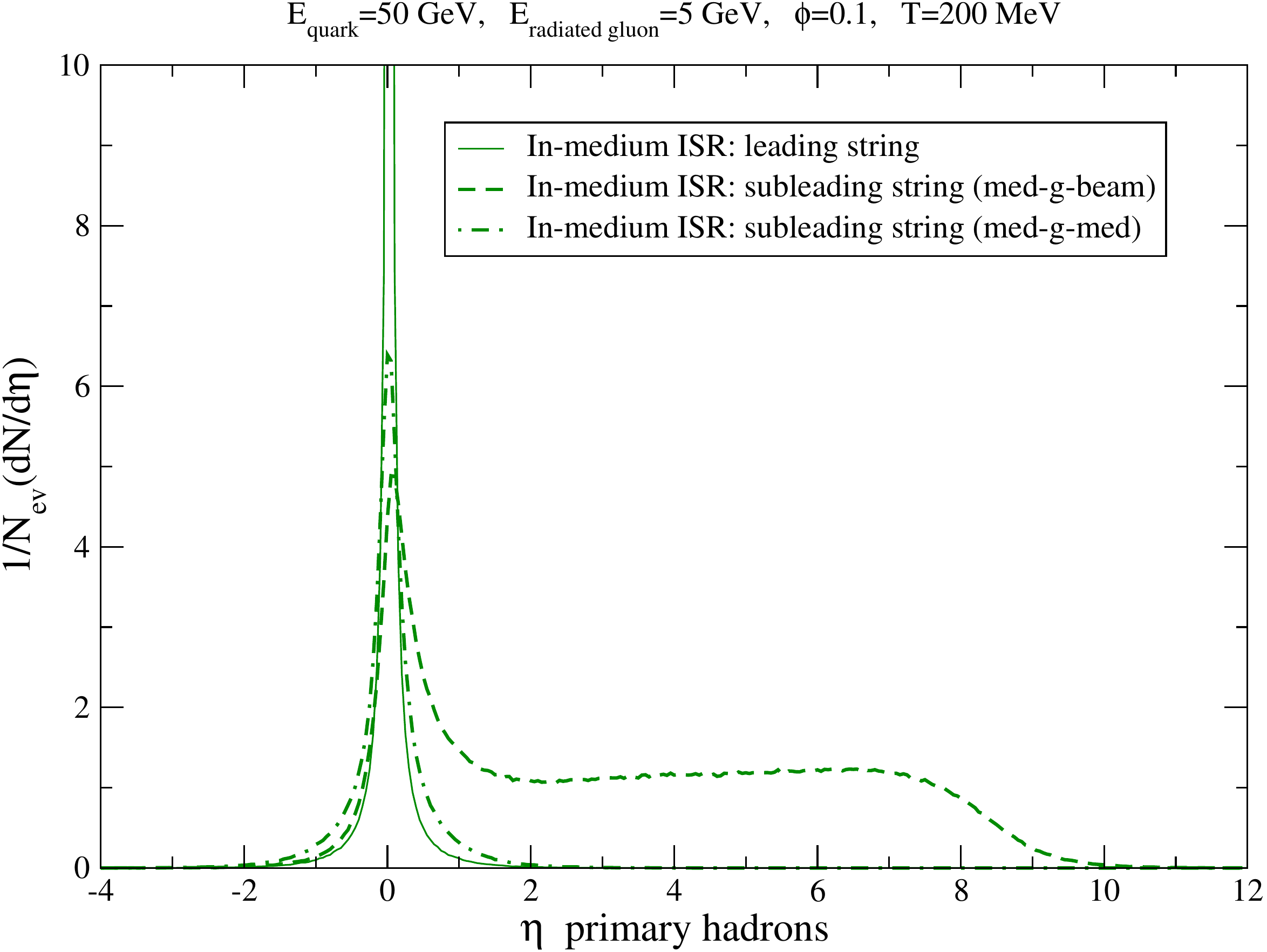}
\caption{The $p_T$ and $\eta$ distributions  of the hadrons from the fragmentation of the Lund strings shown in Fig.~\protect\ref{fig:medshow}. Both the quark and the gluon are emitted at mid-rapidity ($\eta\!=\!0$) at relative angle $\phi\! =\! 0.1$. Left panel: fragmentation pattern in the FSR (in red) and ISR (in green) color channels. Right panel: rapidity distribution of the hadrons in the ISR channel. The sharpest peak around $\eta\!=\!0$ (continuous line) comes from the fragmentation of the leading string. The pattern ``broad peak + plateau'' (dashed line) arises from the fragmentation of the subleading string, connected to the beam remnant (hence the long plateau). Also shown (dot-dashed line) is the case in which both endpoints of the subleading string are attached to a medium particle.} 
\label{fig:FFmed}
\end{center}
\end{figure}
The effects of the modifications of color flow on final hadron distributions are manifest in Fig.~\ref{fig:FFmed}, obtained applying the Lund string-fragmentation scheme implemented in PYTHIA 6.4~\cite{PYTHIA} to typical partonic configurations arising from the interaction with the medium, sharing the same kinematics but having different color connections~\cite{ber2}. The red and green curves in the left panel of Fig.~\ref{fig:FFmed} refer to the FSR and ISR channels shown in Fig.~\ref{fig:medshow}.
ISR is characterized by a depletion of the hard tail of the Fragmentation Function, due to the color decoherence of the emitted gluon, and by an enhancement of soft hadron production from the decay of the subleading string. In the right panel of Fig.~\ref{fig:FFmed} the $\eta$-distribution of hadrons from the fragmentation of the leading and subleading strings in the ISR channel is displayed: the latter, if directly connected to a beam-remnant, would give rise to a long rapidity plateau.

\begin{figure}[!h]
\begin{center}
\includegraphics[clip,width=0.48\textwidth]{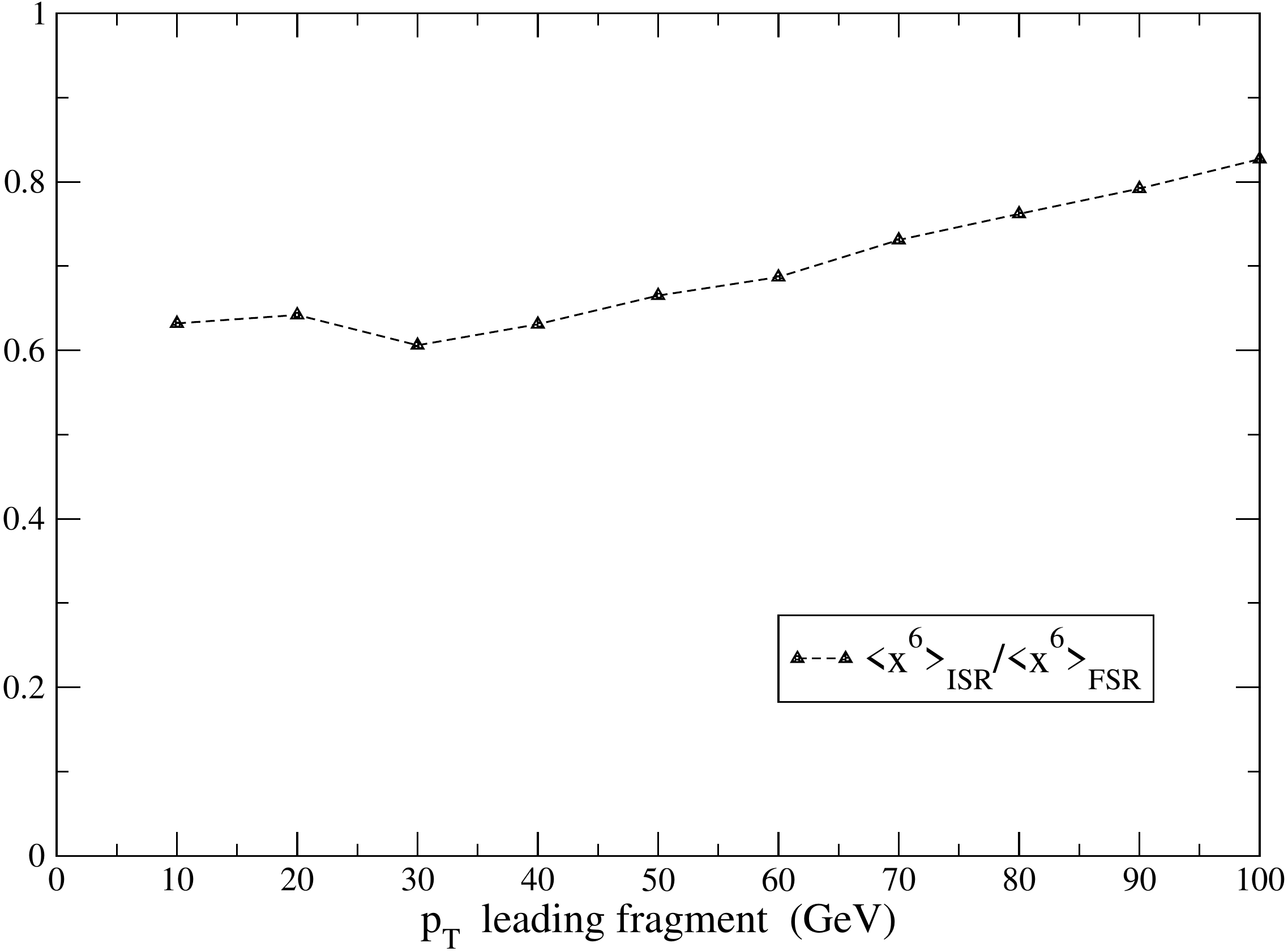}
\caption{The ratio of $6$-th moments of the ISR and FSR contributions to the fragmentation functions of the partonic configurations shown in Fig.~\ref{fig:medshow}. Here, prior to hadronization, the leading quark has transverse momentum $p_T$, and the radiated gluon carries a $k_T\!=\!0.1 p_T$ and is emitted at $\phi = 0.1$. Since in a realistic situation with multiple color exchanges with the medium most of the radiated gluons are color decohered from the leading fragment, the ratio is representative of the effect on the hadron $R_{AA}$ of the modified color connections.} 
\label{fig:moments}
\end{center}
\end{figure}
The softening of the high-momentum tail of the FF in the ISR channel may look mild. However, starting from a steeply falling parton spectrum $\sim 1/p_T^n$, the final hadron distribution $dN^h/dp_T\sim\langle x^{n-1}\rangle/p_T^n$ reflects the higher moments of the FF. In Fig.~\ref{fig:moments} we plot then, as an illustrative example, the ratio of its $6$-th moment in the two different channels: this can provide an estimate of the suppression of the final hadron spectra, purely due to the modification of the color connections in the in-medium parton branchings.
\section{Discussion and conclusions}
It looks of interest to stress the relevance of the color-decoherence effects above discussed in light of the recent experimental findings in jet-quenching analysis~\cite{CMS1,CMS2}. In particular, supplementing the calorimetric measurement of dijet imbalance with tracking information, CMS and ATLAS have established important features of jet fragmentation in heavy-ion collisions.
\begin{itemize}
\item No modification with respect to the $pp$ case was observed for the hard part of the FF (a $p_T$-cut of 4 GeV was used by CMS): this finding can be nicely accommodated within our picture, based on the fragmentation of standard Lund strings of reduced energy (emitted gluons being color-decohered due to the interaction with the medium).
\item The missing energy of the unbalanced jet was found to be carried by soft hadrons with a broad angular distribution. Our description actually foresees that the radiated gluons, in the channels in which they are color-decohered from the projectile, are hadronized as part of independent strings, hence contributing to a sizable enhancement of soft particle multiplicity: the mechanism we propose acts at the interface between partonic and hadronic worlds and is due to the medium-modification of color connections of the emitted gluons. On the other hand the wide-angle hadron emission around the jet-cone would probably require a larger amount of rescattering of the radiated gluons, while for simplicity in Refs.~\cite{ber1,ber2} -- in order to illustrate the effects of color flow -- we confined our analysis to the orders $N\!=\!1$ and $N\!=\!2$ in the opacity expansion (i.e. allowing at most two elastic scatterings in the medium). Whether the anti-angular ordering~\cite{yac1,yac2} of medium-induced radiation might be or relevance in order to explain such a wide angular distribution of soft hadrons is an interesting question. Notice then, in the vacuum, angular-ordering was shown to be at the basis of a depletion of soft hadron production in jet fragmentation (hump-backed plateau): the relaxation of such a constraint due to in-medium radiation might also contribute to an enhancement of soft particle multiplicities. 
\end{itemize}

In our analysis we showed how the color-exchange interaction with the medium is at the basis of the decoherence of the radiated gluons and how this can lead to a softening of the jet FF (hence to a suppression of high-$p_T$ hadron spectra) and to an enhanced production of soft particles. No particular hypothesis was done about the nature of the medium itself (beside the possibility of exchanging color with it), so that the question whether similar effects can be observed also in different systems, like in proton-nucleus collisions, may appear legitimate. 
Actually, though considering jets of much lower transverse energies, similar effects were also observed at Fermilab going from $pp$ to $pA$ collisions and becoming stronger for larger $A$~\cite{E557}: angular broadening of soft particle distributions, softening of the jet FF and even a dijet momentum imbalance. Finally notice that, in simulations of $pA$ collisions performed with the HIJING 2.1 event generator~\cite{wang}, a suppression of high-$p_T$ hadron spectra was found going from an independent fragmentation approach to the Lund hadronization scheme.

Let us now go back to our starting point, namely the factorized ansatz in Eq.~(\ref{eq:fact}). We have shown how the standard hadronization schemes were developed (and tuned) to reproduce data in elementary collisions: a situation (as depicted in Fig.~\ref{fig:vacshow}) in which most of the radiated gluons remain color-connected with the leading high-$p_T$ fragment of the shower. Since, on the contrary, gluons radiated due to the interaction with the medium are decohered from the projectile, a naive convolution $D_{\rm med}\!=\!P(\Delta E)\otimes D_{\rm vac}$ would result into a too hard hadron spectrum: reproducing the experimental amount of quenching would require an overestimate of the parton energy loss.
On the other hand accounting for the modification of the color connections in the presence of a medium would allow to accommodate the experimental data with milder values of the transport parameters (like e.g. the widely used parameter $\hat{q}$).



\begin{thebibliography}{00}
\bibitem{PHENIX} PHENIX Collaboration (K. Adcox et al.),
{\it Suppression of Hadrons with Large Transverse Momentum in Central Au+Au Collisions at  $\sqrt{s_{NN}} = 130$ GeV},
Phys. Rev. Lett. 88 (2002), 022301.
\bibitem{STAR} STAR Collaboration (J. Adams et al.),
{\it Transverse-Momentum and Collision-Energy Dependence of High-pT Hadron Suppression in Au+Au Collisions at Ultrarelativistic Energies}
Phys. Rev. Lett. 91 (2003), 172302. 
\bibitem{ALICE} ALICE Collaboration (K. Aamodt et al.),
{\it {Suppression of Charged Particle Production at Large Transverse Momentum in Central Pb-Pb Collisions at $\sqrt{s_{NN}} = 2.76$ TeV}},
Phys. Lett. B696 (2011), 30.
\bibitem{ATLAS} { ATLAS Collaboration} (G. Aad et al.),
{\it {Observation of a Centrality-Dependent Dijet Asymmetry in Lead-Lead Collisions at $\sqrt{s_{NN}}=2.76$ with the ATLAS Detector at the LHC}},
Phys. Rev. Lett. 105 (2010), 252303.
\bibitem{CMS1} CMS Collaboration (S. Chatrchyan et al.),
{\it Observation and studies of jet quenching in PbPb collisions at $\sqrt{s_{NN}}=2.76$ TeV},
Phys. Rev. C84 (2011), 024906.
\bibitem{CMS2} CMS Collaboration (S. Chatrchyan et al.),
{\it Measurement of jet fragmentation into charged particles in pp and PbPb collisions at $\sqrt{s}_{NN}=2.76$ TeV},
arXiv:1205.5872. 
\bibitem{gui} J. Casalderrey-Solana, J.G. Milhano and U.A. Wiedemann
{\it Jet quenching via jet collimation}
J. Phys. G G38 (2011), 124086.
\bibitem{brick} N. Armesto et al.,  	
{\it Comparison of Jet Quenching Formalisms for a Quark-Gluon Plasma `Brick'},
arXiv:1106.1106.
\bibitem{angord} R.K. Ellis, G. Marchesini and B.R. Webber,
{\it Soft radiation in parton-parton scattering},
Nucl. Phys. B286 (1987), 643.
\bibitem{opal} OPAL collaboration (G. Abbiendi et al.),
{\it Charged particle momentum spectra in e+ e- annihilation at $\sqrt{s}$ = 192-GeV to 209-GeV},
Eur. Phys. J. C27 (2003), 467.
\bibitem{hb} Y.I Azimov et al.,
{\it Humpbacked QCD Plateau in Hadron Spectra}
Z. Phys. C31 (1986) 213.
\bibitem{string} Y.I. Azimov et al.,
{\it The string effect and QCD coherence},
Phys. Lett. B165 (1985), 147.
\bibitem{delphi} DELPHI collaboration (P. Abreu et al.),
{\it Energy dependence of the differences between the quark and gluon jet fragmentation},
Z. Phys. C70 (1996), 179.
\bibitem{CDF} CDF collaboration (F. Abe et al.),
{\it Evidence for color coherence in $p\overline{p}$ collisions at $\sqrt{s}$=1.8 TeV},
Phys. Rev. D50 (1994), 5562.
\bibitem{yac1}
Y.Mehtar-Tani, C.A. Salgado and Konrad Tywoniuk,
{\it Anti-angular ordering of gluon radiation in QCD media}
Phys. Rev. Lett. 106 (2011), 122002
\bibitem{yac2}	
Y. Mehtar-Tani, C.A. Salgado and K. Tywoniuk,
{\it Jets in QCD Media: From Color Coherence to Decoherence}
Phys. Lett. B707 (2012), 156.
\bibitem{iancu} J. Casalderrey-Solana and E. Iancu,
{\it Interference effects in medium-induced gluon radiation},
JHEP 1108 (2011), 015.
\bibitem{ber1} A. Beraudo, J.G. Milhano and U.A. Wiedemann,
{\it Medium-induced color flow softens hadronization}
Phys. Rev. C85 (2012), 031901.
\bibitem{ber2} A. Beraudo, J.G. Milhano and U.A. Wiedemann,
{\it The contribution of medium-modified color flow to jet quenching}
arXiv:1204.4342 (accepted for publication in JHEP).
\bibitem{PYTHIA} T. Sjostrand, S. Mrenna and P.Z. Skands,
{\it PYTHIA 6.4 Physics and Manual}
JHEP 0605 (2006), 026.
\bibitem{E557} E557 collaboration (C. Stewart et al.),
{\it Production of high-pt jets in hadron-nucleus collisions},
Phys. Rev. D42 (1990), 1385.
\bibitem{wang} R. Xu, W.T. Deng and X.N. Wang,
{\it Nuclear modification of high-$p_T$ hadron spectra in p+A collisions at LHC},
arXiv:1205.5019.
\end{thebibliography}







\end{document}